\documentclass[12pt]{iopart}

\makeatletter
\expandafter\let\csname equation*\endcsname\relax
\expandafter\let\csname endequation*\endcsname\relax
\makeatother
\usepackage{amsmath,amssymb}
\makeatletter
\let\save@mathaccent\mathaccent
\newcommand*\if@single[3]{%
  \setbox0\hbox{${\mathaccent"0362{#1}}^H$}%
  \setbox2\hbox{${\mathaccent"0362{\kern0pt#1}}^H$}%
  \ifdim\ht0=\ht2 #3\else #2\fi
  }
\newcommand*\rel@kern[1]{\kern#1\dimexpr\macc@kerna}
\newcommand*\widebar[1]{\@ifnextchar^{{\wide@bar{#1}{0}}}{\wide@bar{#1}{1}}}
\newcommand*\wide@bar[2]{\if@single{#1}{\wide@bar@{#1}{#2}{1}}{\wide@bar@{#1}{#2}{2}}}
\newcommand*\wide@bar@[3]{%
  \begingroup
  \def\mathaccent##1##2{%
    \let\mathaccent\save@mathaccent
    \if#32 \let\macc@nucleus\first@char \fi
    \setbox\z@\hbox{$\macc@style{\macc@nucleus}_{}$}%
    \setbox\tw@\hbox{$\macc@style{\macc@nucleus}{}_{}$}%
    \dimen@\wd\tw@
    \advance\dimen@-\wd\z@
    \divide\dimen@ 3
    \@tempdima\wd\tw@
    \advance\@tempdima-\scriptspace
    \divide\@tempdima 10
    \advance\dimen@-\@tempdima
    \ifdim\dimen@>\z@ \dimen@0pt\fi
    \rel@kern{0.6}\kern-\dimen@
    \if#31
      \overline{\rel@kern{-0.6}\kern\dimen@\macc@nucleus\rel@kern{0.4}\kern\dimen@}%
      \advance\dimen@0.4\dimexpr\macc@kerna
      \let\final@kern#2%
      \ifdim\dimen@<\z@ \let\final@kern1\fi
      \if\final@kern1 \kern-\dimen@\fi
    \else
      \overline{\rel@kern{-0.6}\kern\dimen@#1}%
    \fi
  }%
  \macc@depth\@ne
  \let\math@bgroup\@empty \let\math@egroup\macc@set@skewchar
  \mathsurround\z@ \frozen@everymath{\mathgroup\macc@group\relax}%
  \macc@set@skewchar\relax
  \let\mathaccentV\macc@nested@a
  \if#31
    \macc@nested@a\relax111{#1}%
  \else
    \def\gobble@till@marker##1\endmarker{}%
    \futurelet\first@char\gobble@till@marker#1\endmarker
    \ifcat\noexpand\first@char A\else
      \def\first@char{}%
    \fi
    \macc@nested@a\relax111{\first@char}%
  \fi
  \endgroup
}
\makeatother
\usepackage[
    backend=biber, 
    bibstyle=ieee, 
    citestyle=numeric-comp]{biblatex} 
\usepackage{geometry} 
\usepackage{graphicx}
\usepackage{hyperref} \usepackage[nameinlink]{cleveref}
\usepackage{placeins} 
\usepackage{subcaption}  
\usepackage{tabularx}
\usepackage{url}
\usepackage{upgreek}
\usepackage{xcolor}
\definecolor{blue_hyperref}{HTML}{0D54A6}
\hypersetup{
    colorlinks=true,
    linkcolor=blue_hyperref,
    citecolor=blue_hyperref,
    filecolor=blue_hyperref,
    urlcolor=blue_hyperref,
    pdftitle={Friedel et al. - Magneto-optical signal from Co2Mn-based Heusler thin films in MOKE and BLS},
}
\newcommand{\CMSi}{Co$_2$MnSi}
\newcommand{\CMAl}{Co$_2$MnAl}
\newcommand{\CMGa}{Co$_2$MnGa}
\newcommand{\CMGe}{Co$_2$MnGe}
\newcommand{\CMSn}{Co$_2$MnSn}
\newcommand{\CMX}{Co$_2$Mn\textit{X}}
\newcommand{\theCMFS}{Co$_2$Mn$_{0.6}$Fe$_{0.4}$Si}
\newcommand{\CMFS}{Co$_2$Mn$_x$Fe$_{1-x}$Si}
\newcommand{\CMAS}{Co$_2$MnAl$_x$Si$_{1-x}$} 
 
\newcommand{\CMGxGx}{Co$_2$MnGa$_{x}$Ge$_{1-x}$}

\usepackage{soul} 

\begin{document}

\title[Magneto-optical signal from \texorpdfstring{Co$_2$Mn-based}{Co2Mn-based} Heusler thin films in MOKE and BLS]{Magneto-optical signal from \texorpdfstring{Co$_2$Mn-based}{Co2Mn-based} Heusler thin films in MOKE and BLS}

\author{Anna Maria Friedel$^{1,2}$\footnote{Present address: Institute of Physics, Czech Academy of Sciences, Prague, 162 00 Praha 6, Czech Republic.}, Nicolas Fermon$^{3}$\footnote{Present address: Laboratoire d'Ecologie Alpine, Université Grenoble Alpes, CNRS, Grenoble 38000, France.}, Tobias Böttcher$^{1}$, Sébastien Petit-Watelot$^{2}$, Stéphane Andrieu$^{2}$, Philipp Pirro$^{1}$}

\address{$^1$ Fachbereich Physik and Landesforschungszentrum OPTIMAS, Rheinland-Pf\"alzische Technische Universit\"at Kaiserslautern-Landau, 67663 Kaiserslautern, Germany}
\address{$^2$ Institut Jean Lamour, UMR CNRS 7198, Université de Lorraine, 54000 Nancy, France}
\address{$^3$ \'Ecole Normale Sup\'erieure Paris-Saclay, 91190 Gif-sur-Yvette, France}
\ead{friedel@rptu.de}
\vspace{10pt}
\begin{indented}
\item[]23 October 2026
\end{indented}

\begin{abstract}
Co$_2$Mn-based Heusler compounds offer a versatile, composition-tunable platform for magnonics and spintronics. Among them, the half-metallic \CMSi{} is of particular interest for magnonics owing to its ultralow Gilbert damping, yet its weak magneto-optical response in the visible challenges optical probing such as Brillouin light scattering (BLS). We study the magneto-optical response of epitaxial \CMX{} films (\textit{X} = \{Al$_x$Si$_{1-x}$, Ga$_x$Ge$_{1-x}$, Sn\}) by magneto-optical Kerr effect (MOKE) spectroscopy and BLS. Angle-resolved MOKE resolves a significant, wavelength-dependent quadratic MOKE (QMOKE) only for \CMSi{}, whereas \CMAl{}, \CMGa{} and \CMSn{} respond dominantly linearly. Comparing BLS intensities of thermal magnons at two wavelengths, \CMSi{} gives the weakest signal at $532\,\mathrm{nm}$ yet among the strongest at $457\,\mathrm{nm}$, tracking the spectral dependence of its Kerr angle. These results emphasise the relation between the two magneto-optical techniques, guiding the choice of probing wavelength for Co$_2$Mn-based Heusler compounds, and provide a transferable approach for other magnetic materials.
\end{abstract}
%

%
%
%
%
\section{Introduction}\label{sec:intro}
Magneto-optical experiments form a core group of experimental techniques in magnetism, in particular for the study of magnetisation dynamics. Contrary to typical inductive methods \supercite{vlaminck2008CurrentInduced, ciubotaru2016All}, the magneto-optical probing brings the advantage of principally not requiring sample patterning. Moreover, the versatile range of magneto-optical techniques spans from table-top Kerr microscopy \supercite{soldatov2017Advances} to sophisticated methods like super-Nyquist sampling magneto-optical Kerr effect (SNS-MOKE) experiments \supercite{dreyer2021Spinwave}, Brillouin light scattering (BLS) spectroscopy \supercite{sebastian2015Microfocused}, or even involving synchrotron facilities \supercite{forster2019Nanoscale}. 

However, a common challenge for magneto-optical probing is the requirement of sufficient magneto-optical signal intensities \supercite{lozovski2024Plasmonenhanced, huang2026Advancing}, increasingly driving engineering efforts for signal enhancement \supercite{maksymov2015MagnetoPlasmonics, yu2019Plasmonenhanced}. Intrinsically, the decisive role is played by the band structure of the material under investigation defining the magneto-optical response for a specific probing setup (light wavelength and geometry). And while the band structure can be engineered to enhance magneto-optical signal intensities e.g. by doping, see for instance YIG vs. Bi:YIG \supercite{wittekoek1975Magnetooptic}, this generally comes at the expense of various other material parameters. This is for instance the case in Co$_2$Mn-based Heusler compounds: The family contains several half-metallic candidates, where particularly \CMSi{} has recently been established as a realistic material choice for nano-fabricated half-metal magnonic devices \supercite{friedel2026Epitaxial, cordova2026Spinpolarization}. Recent advances in epitaxial growth \supercite{guillemard2020Issues} finally allow for reliable 100\% spin polarisation of the electronic current even in patterned \CMSi{} devices \supercite{cordova2026Spinpolarization}, combined with an associated ultralow Gilbert damping maintained down to cryogenic temperatures \supercite{guillemard2019Ultralow, demelo2021Unveiling}. Ensuring this long-term sought-after combination at the intersection of magnonics and spintronics \supercite{hillebrands2006Highspin} has refuelled research interest within the magnonics community \supercite{mantion2024Reconfigurable, serha2026Magnetic}, further corroborated by the reported nanofabrication robustness down to device sizes of $50\,\mathrm{nm}$ \supercite{friedel2026Epitaxial}. Yet, \CMSi{} is known to present poor magneto-optical signal in the visible light range as described by its vanishing Kerr angle \supercite{silber2020Scaling}. This difficult magneto-optical accessibility had previously directed the magnonics community to focus on the promising compound \theCMFS{} instead \supercite{sebastian2012Lowdamping, sebastian2013Nonlinear, pirro2014NonGilbertdamping, sebastian2015Alloptical, sebastian2016Chapter}. And while the tuning in quaternary Heusler compounds such as \CMFS{} or \CMAS{} indeed allows for improved magneto-optical signal intensities \supercite{friedel2021Thin}, the change in stoichiometry simultaneously tunes key parameters and is now known to compromise spin polarisation, Gilbert damping and exchange constant \supercite{kubota2009Structure, guillemard2020Engineering}.  

Therefore, a more desirable first approach for the magneto-optical signal improvement is the optimised choice of probing wavelength and geometry. With the majority of magneto-optical probing techniques being directly related to the magneto-optical Kerr effect (MOKE) \supercite{soldatov2017Advances, dreyer2021Spinwave, huang2026Advancing, hamrle2010Analytical}, this requirement can be generally reformulated as the selection of probing wavelengths associated with a non-zero, ideally maximised complex Kerr angle. MOKE spectroscopy is a powerful tool to quantify the wavelength-dependence \supercite{silber2020Scaling}, and recent reports on the higher order MOKE effects additionally suggest a consideration of the probing geometry particularly for samples with a monocrystalline, cubic lattice structure \supercite{gaerner2024Cubic, silber2026Cubicinmagnetizationa, gaerner2026Cubic}. Indeed, L2\textsubscript{1}-ordered Heusler compounds like Co\textsubscript{2}FeSi, \CMGe{} and \CMSi{} are known to exhibit significant quadratic magneto-optical Kerr effect (QMOKE) contributions in addition to the linear (linMOKE) effect \supercite{hamrle2007Huge, gaier2008Influence, muduli2008Composition}, whereby the strong dependence on the spin-orbit coupling \supercite{hamrle2007Ion} reportedly yields a sensitivity on composition deviation \supercite{muduli2008Composition} or crystallographic ordering \supercite{silber2020Scaling}. However, most studies focus on a single compound, raising the question on the comparability between the different related compounds in the \CMX{} family, particularly in view of the reported tunability via quaternary compounds like \CMAS{}. Moreover, with the analytically established link between complex Kerr angle and BLS intensity \supercite{hamrle2010Analytical}, the question arises whether MOKE spectrometry can directly guide the optimal wavelength choice for BLS experiments.  

In this work, we therefore address three questions on the magneto-optical response of \CMX{} Heusler compounds: (i) How do the higher-order, quadratic MOKE and its directional anisotropy compare across the \CMX{} family? (ii) How does this response depend on the probing wavelength, and is it a robust material property? (iii) Can MOKE spectroscopy directly guide the choice of BLS probing wavelength, i.e. does the analytical link between MOKE and BLS intensity\supercite{hamrle2010Analytical} hold in practice? Performing MOKE spectroscopy on the epitaxial \CMX{} films, we compare the magneto-crystalline anisotropy and the impact on the second-order magneto-optical anisotropy at $\lambda = 636\,\mathrm{nm}$ and thereby resolve a significant QMOKE among the ternary compounds only for L2\textsubscript{1}-ordered \CMSi{}. We study the wavelength dependence across $\lambda = 550\,\mathrm{nm} - 900\,\mathrm{nm}$, where the Kerr angle of \CMSi{} is strongly dispersive and reproduces the reference spectra of Silber et al.\supercite{silber2020Scaling}, whereas B2-disordered \CMAl{} remains nearly linear throughout the probed range. Finally, we compare the BLS intensity from thermally excited magnon spectra across the \CMX{} series for two different wavelengths $\lambda_\mathrm{BLS} = 457\,\mathrm{nm}$ and $532\,\mathrm{nm}$. Hereby, \CMSi{} gives the weakest BLS signal of the series at $532\,\mathrm{nm}$ but among the strongest at $457\,\mathrm{nm}$, tracking the spectral dependence of the Kerr angle. These experimental results are consistent with the analytical link between MOKE and BLS, showing that MOKE spectroscopy can guide the choice of BLS probing wavelength. Concomitantly, the research in this work suggests taking the optical anisotropies of the higher-order MOKE contributions into account when planning future related magneto-optical measurements.

\section{Samples and methods}\label{sec:methods}

\subsection{Sample series and material characteristics}\label{sec:methods/samples}
The \CMX{} (\textit{X} = \{Al$_x$Si$_{1-x}$, Ga$_x$Ge$_{1-x}$, Sn\}) sample series investigated in this study consists of epitaxial \CMX{} Heusler thin films grown on MgO(001) substrates by molecular beam epitaxy in the framework of a previous study by Guillemard \textit{et al.} \supercite{guillemard2019Ultralow, guillemard2019Halfmetal, guillemard2020Engineering, guillemard2020Issues}, where the structural and magnetic properties of the films are reported in detail. All \CMX{} were grown on an MgO buffer layer and capped with an Au capping layer to prevent oxidation, and the film normal of all \CMX{} films is [001] \supercite{guillemard2020Issues}. The chemical ordering in the \CMX{} films of the series range from full B2 disorder in the \CMAl{} film to full L2$_1$ order in the \CMSi{} film, resulting in a spin polarisation between $63\,\%$ (\CMAl{}) and $100\,\%$ (\CMSi{}) \supercite{guillemard2020Engineering}. An overview on the relevant material parameters is given in \Cref{table:1}. 

\begin{table}
\centering
\caption{Sample and material parameters of the \CMX{} (\textit{X} = \{Al$_x$Si$_{1-x}$, Ga$_x$Ge$_{1-x}$, Sn\}) sample series investigated in this work, all originally reported in the publications by Guillemard et al. \supercite{guillemard2019Ultralow, guillemard2019Halfmetal, guillemard2020Engineering, guillemard2020Issues}: 
\CMX{} film thicknesses $t_{\mathrm{Co2Mn}X}$, 
effective magnetisation $M_\mathrm{eff}$ from ferromagnetic resonance spectroscopy (FMR) measurements, 
predominant chemical ordering from high angle annular dark field scanning transmission electron microscopy (HAADF-STEM) measurements, 
spin polarisation $P_{\uparrow\downarrow}$ from spin resolved photo emission spectroscopy (SR-PES),
and Gilbert damping parameter $\alpha$ from FMR measurements.}
\begin{tabularx}{\textwidth}{
    >{\raggedright\arraybackslash}p{35mm} |
    >{\raggedright\arraybackslash}X 
    >{\raggedright\arraybackslash}X 
    >{\raggedright\arraybackslash}p{25mm} 
    >{\raggedright\arraybackslash}X 
    >{\raggedright\arraybackslash}X 
    } 
    Compound 
    & $t_{\mathrm{Co2Mn}X}$ 
    & $M_\mathrm{eff}$ 
    & ordering
    & $P_{\uparrow\downarrow}$
    & $\alpha$\\
    %
    %
    & (nm) 
    & ($\mathrm{kA}/\mathrm{m}$) 
    & 
    & (\%)
    & ($\times10^{-3}$) \\
    %
    %
    \hline
    %
    %
    \CMAl{}   
    & 17 \supercite{guillemard2019Halfmetal} 
    & \phantom{1}835 \supercite{guillemard2020Engineering} 
    & B2 \supercite{guillemard2019Ultralow, guillemard2020Engineering}
    & \phantom{1}63 \supercite{guillemard2020Engineering}
    & 1.10 \supercite{guillemard2020Engineering}
    \\
    %
    %
    Co$_{2}$MnAl$_{0.75}$Si$_{0.25}$  
    & 20 \supercite{guillemard2019Halfmetal} 
    & \phantom{1}947 \supercite{guillemard2020Engineering}
    & B2 + L2\textsubscript{1} \supercite{guillemard2020Engineering}
    & \phantom{1}70 \supercite{guillemard2020Engineering}
    & 1.00 \supercite{guillemard2020Engineering}
    \\
    %
    %
    \CMSi{}   
    & 17 \supercite{guillemard2019Halfmetal} 
    & 1049 \supercite{guillemard2019Ultralow} 
    & L2\textsubscript{1} \supercite{guillemard2019Ultralow, guillemard2020Engineering}
    & 100 \supercite{guillemard2019Ultralow}
    & 0.46 \supercite{guillemard2019Ultralow,guillemard2020Engineering}
    \\
    %
    %
    \CMGa{}   
    & 21 \supercite{guillemard2019Halfmetal} 
    & 1042 \supercite{guillemard2019Ultralow} 
    & L2\textsubscript{1} \supercite{guillemard2019Ultralow}
    & 100 \supercite{guillemard2019Ultralow}
    & 2.00 \supercite{guillemard2019Ultralow}
    \\
    %
    %
    Co$_{2}$MnGa$_{0.75}$Ge$_{0.25}$  
    & 21 \supercite{guillemard2019Halfmetal} 
    & \phantom{1}850 \supercite{guillemard2019Halfmetal} 
    & 
    & 
    & 1.00 \supercite{guillemard2019Halfmetal}\\
    %
    %
    Co$_{2}$MnGa$_{0.5}$Ge$_{0.5}$  
    & 21 \supercite{guillemard2019Halfmetal} 
    & \phantom{1}850 \supercite{guillemard2019Halfmetal} 
    & 
    & 
    & 1.00 \supercite{guillemard2019Halfmetal}\\
    %
    %
    \CMGe{}   
    & 21 \supercite{guillemard2019Halfmetal} 
    & \phantom{1}975 \supercite{guillemard2019Ultralow} 
    & L2\textsubscript{1} \supercite{guillemard2019Ultralow}
    & 100 \supercite{guillemard2019Ultralow}
    & 0.53 \supercite{guillemard2019Ultralow}
    \\
    %
    %
    \CMSn{}   
    & 25 \supercite{guillemard2019Halfmetal} 
    & \phantom{1}961 \supercite{guillemard2019Ultralow} 
    & X+L2\textsubscript{1} \supercite{guillemard2019Ultralow, guillemard2019Halfmetal, guillemard2020Issues}
    & \phantom{1}81 \supercite{guillemard2019Ultralow}
    & 0.90 \supercite{guillemard2019Ultralow}
    \\
\end{tabularx}\label{table:1}
\end{table}

\subsection{Magneto-optical Kerr effect (MOKE) spectroscopy}\label{sec:methods/MOKE}

\begin{figure}[ht!]
  \centering
  \includegraphics{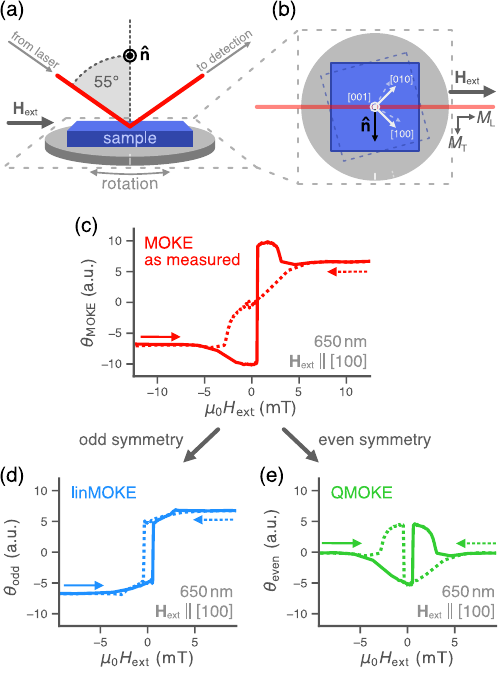}
  \caption[MOKE detection and symmetry separation]{
    \textbf{MOKE detection and symmetry separation} 
    \textbf{(a)} Schematic of the MOKE setup, with the sample positioned on a rotatable plate in an external field parallel to the incidence plane of light defined by the normal $\hat{\mathbf{n}}$.  
    \textbf{(b)} Top view on the sample plane with the crystal coordinate system of the \CMX{} compounds. 
    \textbf{(c)} MOKE hysteresis loop as measured for \CMSi{} at $\lambda = 650\,\mathrm{nm}$ and $\mathbf{H}_\mathrm{ext}\parallel \left[100\right]$. In a post-processing step, the signal is separated into \textbf{(d)} anti-symmetric part attributed to the linMOKE and \textbf{(e)} symmetric part attributed to the QMOKE. 
  }
  \label{fig:1_MOKE_separation}

  \begin{subfigure}{0pt}
    \phantomcaption
    \label{fig:1a}
  \end{subfigure}
  \begin{subfigure}{0pt}
    \phantomcaption
    \label{fig:1b}
  \end{subfigure}
  \begin{subfigure}{0pt}
    \phantomcaption
    \label{fig:1c}
  \end{subfigure}
  \begin{subfigure}{0pt}
    \phantomcaption
    \label{fig:1d}
  \end{subfigure}
  \begin{subfigure}{0pt}
    \phantomcaption
    \label{fig:1e}
  \end{subfigure}
\end{figure}

Magneto-optical Kerr effect (MOKE) spectroscopy was carried out on a standard longitudinal MOKE (LMOKE) setup with the external magnetic field $\mathbf{H}_\mathrm{ext}$ applied in the film plane and perpendicular to the normal $\hat{\mathbf{n}}$ on the incidence plane of light. The light is provided by a NKT photonics SuperK EXTREME supercontinuum laser, providing output in a wavelength range of $\lambda_\mathrm{MOKE} \approx 550\,\mathrm{nm} - 900\,\mathrm{nm}$, and mounted at an angle of incidence of $55^\circ$ to the film normal as illustrated in \Cref{fig:1a}. The sample is positioned on a rotatable plate centred in the magnetic field, allowing for a rotational study of the LMOKE as a function of the crystallographic orientation as shown in \Cref{fig:1b}. All measurements were recorded as standard MOKE hysteresis loops, using s-polarised light and carried out at room temperature. 

For the separation of the linear MOKE (linMOKE) and quadratic MOKE (QMOKE) contributions, multiple methods have been proposed in recent studies: the ROTMOKE method \supercite{mattheis1999Separation, mattheis1999Determination}, the 8 field method \supercite{hamrle2007Huge, postava2002Anisotropy, silber2018Quadratic}, the separation based on field-dependent symmetry considerations \supercite{mewes2004Separation, hamrle2007Huge, gaier2008Influence}, and the rotating field method \supercite{liang2015Separation}. In this work, the symmetry-based separation approach \supercite{mewes2004Separation, hamrle2007Huge, gaier2008Influence} was performed in a post-processing step and an example is illustrated in \Cref{fig:1_MOKE_separation}: The measured signal $\theta_\mathrm{MOKE}$ as shown in \Cref{fig:1c} is separated into the odd (anti-symmetric) and even (symmetric) parts 
\begin{align}
    \theta_\mathrm{odd} &= \frac{\theta_\mathrm{inc}(H)-\theta_\mathrm{dec}(-H)}{2} \quad \rightarrow \quad \theta_\mathrm{linMOKE}\\
    \theta_\mathrm{even} &= \frac{\theta_\mathrm{inc}(H)+\theta_\mathrm{dec}(-H)}{2} \quad \rightarrow \quad \theta_\mathrm{QMOKE}
\end{align}
as shown in \Cref{fig:1d} and \Cref{fig:1e} respectively, with $\theta_\mathrm{inc/dec}$ indicating the recorded hysteresis loop branch for increasing and decreasing external field. Assuming negligible higher orders $\mathcal{O}(M^3)$ to the magneto-optical response, $\theta_\mathrm{odd}$ and $\theta_\mathrm{even}$ can be attributed to first and second order effects commonly denominated as linear MOKE (linMOKE) $\theta_\mathrm{linMOKE}$ and quadratic MOKE (QMOKE) $\theta_\mathrm{QMOKE}$ respectively. This is a valid approximation, as the cubic MOKE (CMOKE) contribution is known to vanish  for (001)-oriented films studied under medium angles of incidence \supercite{silber2026Cubicinmagnetizationa}, and indications for significant $\mathcal{O}(M^4)$ contributions have neither been reported to date nor been identified in our study. 
We further note that unlike the 8 field method \supercite{hamrle2007Huge, postava2002Anisotropy}, this simple symmetry-based separation approach does not allow for the quantitative distinction of $\theta_{M_\mathrm{L} M_\mathrm{T}}\sim M_\mathrm{L} M_\mathrm{T}$ (analogous to the planar Hall effect) and $\theta_{M_\mathrm{L}^2 - M_\mathrm{T}^2} \sim M_\mathrm{L}^2 - M_\mathrm{T}^2$ (analogous to the anisotropic magneto-resistance), i.e. the QMOKE contributions with different symmetries in the longitudinal and transverse in-plane magnetisation components $M_\mathrm{L} \perp \hat{\mathbf{n}}$ and  $M_\mathrm{T} \parallel \hat{\mathbf{n}}$ cannot be distinguished in this study.

\subsection{Brillouin light scattering (BLS) spectroscopy}\label{sec:methods/BLS}

Thermally excited spin waves were measured at room temperature using a conventional BLS system mounted in the backscattering geometry. BLS measurements were carried out for two typical laser wavelengths $\lambda_\mathrm{BLS} = 532\,\mathrm{nm}$ and $\lambda_\mathrm{BLS} = 457\,\mathrm{nm}$, and performed in the Damon-Eshbach geometry, i.e. with the external magnetic field $\mathbf{H}_\mathrm{ext}\parallel\hat{\mathbf{n}}$ applied in the film plane along the normal $\hat{\mathbf{n}}$ on the incidence plane of light. The external field was fixed to  $\mu_0 H_\mathrm{ext} = 100\,\mathrm{mT}$ to ensure a saturation of the samples. Measurements were performed at fixed angles of incidence $\varphi=2.5^\circ$ and $10^\circ$, defining the probed in-plane wavevector $k_\parallel = \frac{4\pi \sin\varphi}{\lambda_\mathrm{BLS}}$ of the spin waves. 

For an improved signal-to-noise ratio, multiple spectra were averaged for each sample\footnote{Specifically, we recorded $1\,000$ scans for each field polarity at $532\,\mathrm{nm}$. At $457\,\mathrm{nm}$, we obtained lower signal and hence averaged over $40\,000$ to $60\,000$ scans per sample. Note that as the BLS signal intensity is normalised to the reference signal, the total number of recorded scans is irrelevant to the evaluation in this study.}. The BLS signal intensities were extracted as the area of a Lorentzian function fitted to the peaks in the recorded spectra, and the corresponding errors were determined as the standard errors of the fitted areas averaged over both peaks. The extracted BLS signal intensity was normalised to the respective reference signal in order to exclude any influence of possible instabilities of the output laser power. The reference signal was extracted from the measurements by summing the recorded signal in the frequency range from $-3\,\mathrm{GHz}$ to $3\,\mathrm{GHz}$ and corresponds to a reference beam split off from the probing beam before interaction with the sample. Conveniently, this normalisation also excludes any influence due to a possible drifting of the interferometer performance over the time of the measurements. 

\FloatBarrier
\section{Results and discussion}

\subsection{Magneto-crystalline and magneto-optical anisotropy}

\begin{figure}[p]
  \centering
  \includegraphics{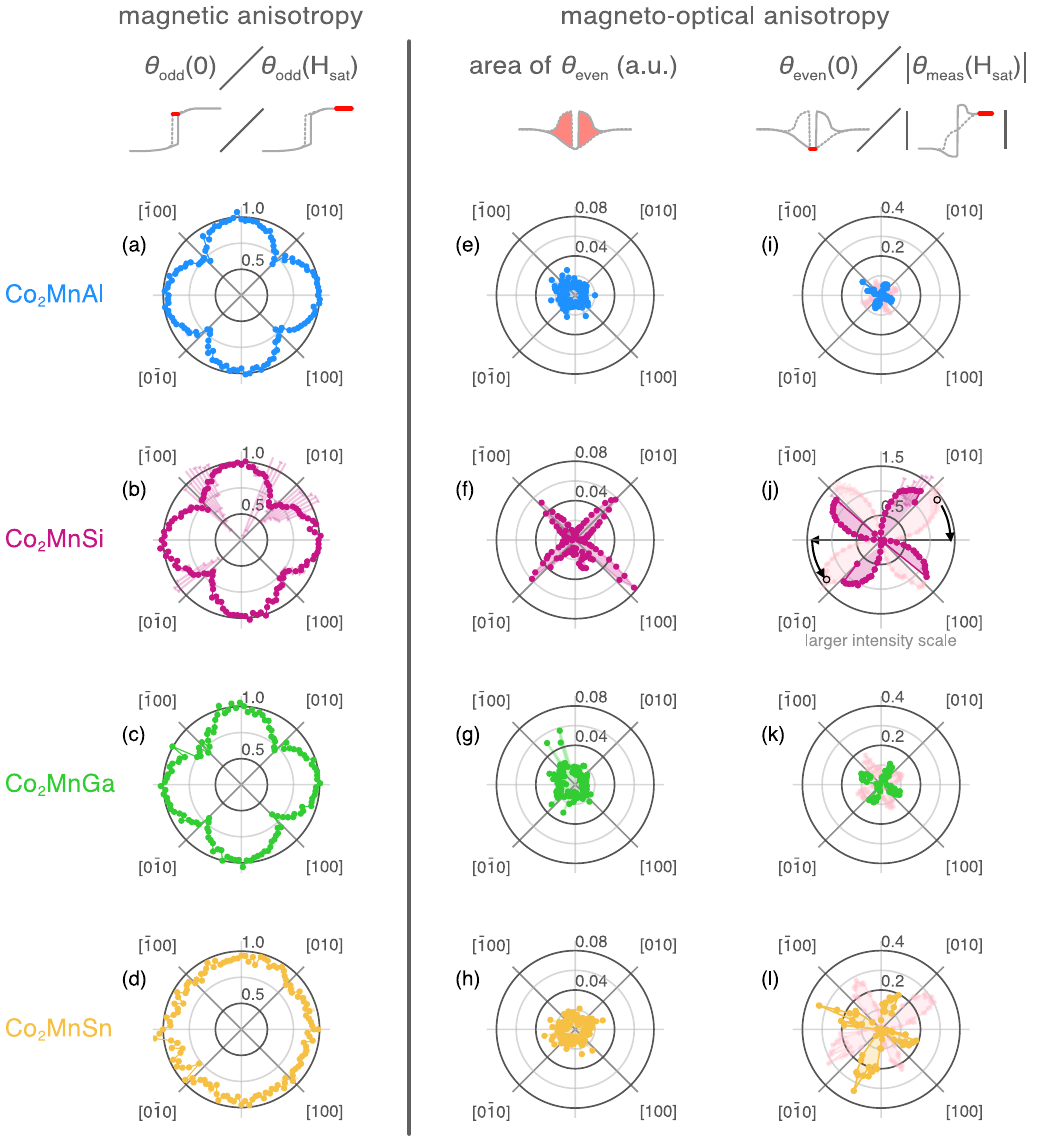}
  \caption[Anisotropy in MOKE]{
    \textbf{Anisotropy in the MOKE signal of various \CMX{} compounds for $\lambda = 636\,\mathrm{nm}$.} 
    (a-d) The magneto-crystalline anisotropy shows in the relative remanence in the odd Kerr rotation $\theta_\mathrm{odd} \left(0\right) / \theta_\mathrm{odd} \left(H_\mathrm{sat}\right)$ attributed to the linMOKE.  
    The magneto-optical anisotropy shows in the integrated QMOKE loop area (e-h) as well as the relative remanent QMOKE $\theta_\mathrm{even} \left(0\right) / | \theta_\mathrm{meas} \left(H_\mathrm{sat}\right) |$ (i-l), which additionally reveals a sign change with the opposite sign reflected by the light pink shade. Note that while the schematic illustrations only mark the relevant points for the decreasing field branch for (a-d) and (i-l), both increasing and decreasing branch were evaluated and averaged.
  }
  \label{fig:2_MOKE_anisotropy}

  \begin{subfigure}{0pt}
    \phantomcaption
    \label{fig:2a}
  \end{subfigure}
  \begin{subfigure}{0pt}
    \phantomcaption
    \label{fig:2b}
  \end{subfigure}
  \begin{subfigure}{0pt}
    \phantomcaption
    \label{fig:2c}
  \end{subfigure}
  \begin{subfigure}{0pt}
    \phantomcaption
    \label{fig:2d}
  \end{subfigure}
  \begin{subfigure}{0pt}
    \phantomcaption
    \label{fig:2e}
  \end{subfigure}
  \begin{subfigure}{0pt}
    \phantomcaption
    \label{fig:2f}
  \end{subfigure}
  \begin{subfigure}{0pt}
    \phantomcaption
    \label{fig:2g}
  \end{subfigure}
  \begin{subfigure}{0pt}
    \phantomcaption
    \label{fig:2h}
  \end{subfigure}
  \begin{subfigure}{0pt}
    \phantomcaption
    \label{fig:2i}
  \end{subfigure}
  \begin{subfigure}{0pt}
    \phantomcaption
    \label{fig:2j}
  \end{subfigure}
  \begin{subfigure}{0pt}
    \phantomcaption
    \label{fig:2k}
  \end{subfigure}
  \begin{subfigure}{0pt}
    \phantomcaption
    \label{fig:2l}
  \end{subfigure}
\end{figure}

We first investigate the anisotropy of the MOKE response for the ternary Heusler compounds \CMAl{}, \CMSi{}, \CMGa{} and \CMSn{}. To this aim, a full rotational scan is performed for each sample at a fixed wavelength of $\lambda = 636\,\mathrm{nm}$, the results of which are presented in \Cref{fig:2_MOKE_anisotropy}. 

To verify the magneto-crystalline anisotropy, we extract the signal at remanence in the odd Kerr rotation $|\theta_\mathrm{odd}\left(0\right)|$ and normalise by the averaged saturation signal $|\theta_\mathrm{odd}\left(H_\mathrm{sat}\right)|=|\theta_\mathrm{odd}\left(|H|\geq 20\,\mathrm{mT}\right)|$, as is shown in \Cref{fig:2a,fig:2b,fig:2c,fig:2d}. For \CMAl{}, \CMSi{} and \CMGa{}, a cubic anisotropy is observed in \Cref{fig:2a,fig:2b,fig:2c} with hard axes along the principal lattice directions $\langle100\rangle$ along which a drop to $M_\mathrm{remanence}\approx 70\% M_\mathrm{sat}$ is obtained, whereas the \CMSn{} film shows an isotropic behaviour in \Cref{fig:2d}. This is in agreement with vibrating sample magnetometry results obtained from these films in the aforementioned precedent work \supercite{guillemard2019Halfmetal}, presenting equally a cubic anisotropy with a drop to $M_\mathrm{remanence}\approx 70\% M_\mathrm{sat}$ along the $\langle100\rangle$ directions for \CMAl{}, \CMSi{} and \CMGa{}, and an insignificant anisotropy in \CMSn{}. 

Contrary to the isotropic linear MOKE for (001)-oriented cubic crystals, the higher order contributions can present a directional dependence on $[hkl] \parallel \mathbf{H}_\mathrm{ext} \perp \hat{\mathbf{n}}$. This magneto-optical anisotropy is observed in the extracted even symmetry part of the Kerr rotation $\theta_\mathrm{even}$ associated to the QMOKE in this study and showcased by the integrated area of the $\theta_\mathrm{even}\left(H\right)$ loops presented in \Cref{fig:2e,fig:2f,fig:2g,fig:2h}. The most significant magneto-optical anisotropy effect for this probing wavelength $\lambda = 636\,\mathrm{nm}$ is observed on \CMSi{} in \Cref{fig:2f} with a clear directional dependence and a maximal effect along the principal lattice directions $\langle100\rangle$. 

For an intuitive representation of the significance of this higher order effect in comparison to the total measured MOKE signal, the even Kerr rotation in remanence $\theta_\mathrm{even}\left(0\right)$ is extracted and normalised by the absolute value of the directly measured saturation signal $|\theta_\mathrm{meas}\left(H_\mathrm{sat}\right)|=|\theta_\mathrm{meas}\left(|H|\geq 20\,\mathrm{mT}\right)|$, the results of which are presented in \Cref{fig:2i,fig:2j,fig:2k,fig:2l}. This reveals a sign change of $\theta_\mathrm{even}\left(0\right)$ upon crossing of a high symmetry direction $\langle100\rangle$ or $\langle110\rangle$, which reflects the rotation towards the nearest easy axis before flipping for a full magnetisation reversal as is schematically indicated by the arrows in \Cref{fig:2j}. While this sign change is observed on all compounds, the normalised even Kerr rotation in \Cref{fig:2i,fig:2j,fig:2k,fig:2l} highlights the insignificance of the effect in relation to the total measured signal for \CMAl{} and \CMGa{} in \Cref{fig:2i,fig:2k}, where $\theta_\mathrm{even}\left(0\right)$ amounts to only $\sim10\%$ of the total signal at $\lambda = 636\,\mathrm{nm}$. 
A marginally larger relative effect is observed for \CMSn{} in \Cref{fig:2l}. We note that \CMSn{} is the only compound in the series with a mixture of inverse Heusler order (X ordering) and full Heusler structure (L2\textsubscript{1} ordering) \cite{guillemard2020Issues}, as listed in \Cref{table:1}, whereas the \CMAl{} film is fully B2-disordered and only \CMGa{} and \CMSi{} are fully L2\textsubscript{1}-ordered. 
The symmetry imposed by the crystal ordering can strongly affect the higher order MOKE contributions, as is known for \CMSi{}, where the effect scales with increased L2\textsubscript{1}-ordering \supercite{wolf2011Quadratic, silber2020Scaling}, consistent with the significant second order contribution up to $\sim150\%$ in \Cref{fig:2j}. Consequently, the comparison across compounds in \Cref{fig:2_MOKE_anisotropy} opens the question on the significance of the ordering-imposed symmetries in relation to other factors affecting the underlying band structure. Yet, given the reported strong wavelength dependence at least in \CMSi{}, a clear conclusion cannot be drawn from the single-wavelength study in \Cref{fig:2_MOKE_anisotropy}.

\FloatBarrier

\subsection{Wavelength dependence of linMOKE and QMOKE for \texorpdfstring{\CMSi{}}{Co2MnSi} and \texorpdfstring{\CMAl{}}{Co2MnAl}}

\begin{figure}[ht!]
  \centering
  \includegraphics{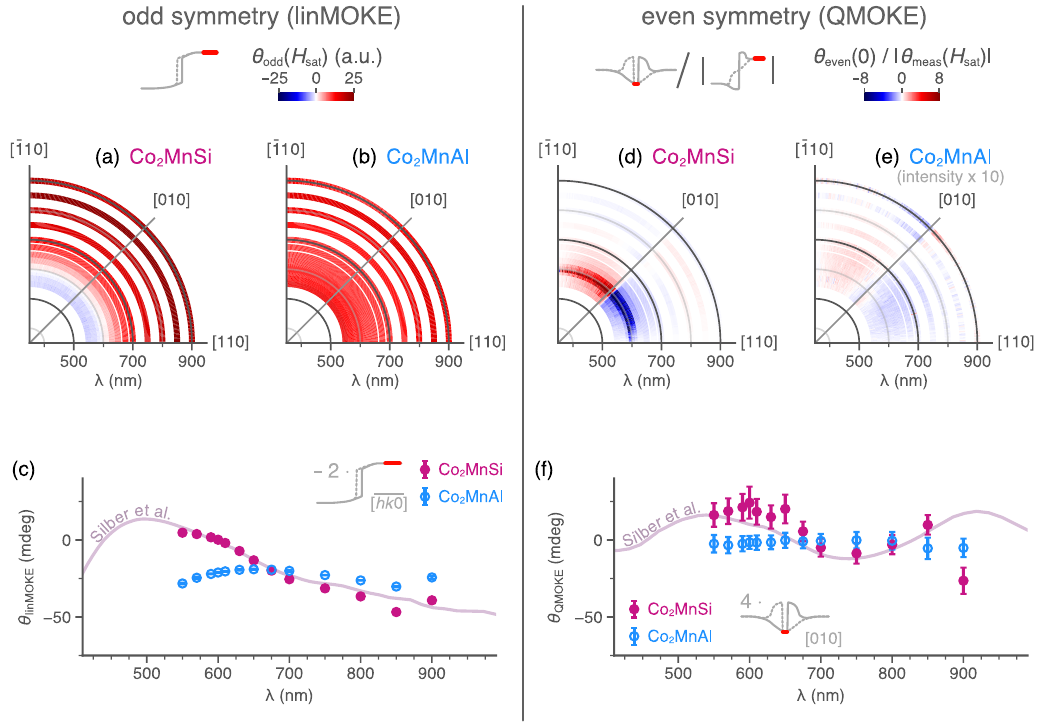}
  \caption[Wavelength dependence of the linear and quadratic MOKE effect in \CMSi{} and \CMAl{}.]{
    \textbf{Wavelength dependence of the linear and quadratic MOKE effect in \CMSi{} and \CMAl{}.} (a, b) The linMOKE rotation in saturation $\theta_\mathrm{odd}\left(H_\mathrm{sat}\right)$ as a function of the crystal orientation $[hkl]\parallel \mathbf{H}_\mathrm{ext} \perp \hat{\mathbf{n}}$ for \CMSi{} and \CMAl{}. (c) The mean value over all directions is compared to the reference study for \CMSi{} by Silber \textit{et al.} \supercite{silber2020Scaling, silber2020Longitudinal}, where the empirical scaling factor $-2$ reflects the opposite coordinate convention, approximately double \CMSi{} layer thickness and related differences between the two studies. 
    (d, e) The even QMOKE rotation in remanence $\theta_\mathrm{even}\left(0\right)$ in relation to the saturation amplitude of the measured loops $\theta_\mathrm{meas}\left(H_\mathrm{sat}\right)$. A significant QMOKE contribution is only detected for \CMSi{}. (f) The wavelength dependence along the $[010]$ direction is compared to the reference study for \CMSi{} by Silber \textit{et al.} \supercite{silber2020Scaling, silber2020Longitudinal}, after applying the empirical scaling factor $4$.
  }
  \label{fig:3_wavelength}

  \begin{subfigure}{0pt}
    \phantomcaption
    \label{fig:3a}
  \end{subfigure}
  \begin{subfigure}{0pt}
    \phantomcaption
    \label{fig:3b}
  \end{subfigure}
  \begin{subfigure}{0pt}
    \phantomcaption
    \label{fig:3c}
  \end{subfigure}
  \begin{subfigure}{0pt}
    \phantomcaption
    \label{fig:3d}
  \end{subfigure}
  \begin{subfigure}{0pt}
    \phantomcaption
    \label{fig:3e}
  \end{subfigure}
  \begin{subfigure}{0pt}
    \phantomcaption
    \label{fig:3f}
  \end{subfigure}
\end{figure}

In view of the application interest in quaternary \CMAS{} compounds for their tunable structural as well as electronic properties \supercite{kubota2009Structure, guillemard2020Engineering}, we select the two associated ternary compounds \CMSi{} and \CMAl{} for an in-depth wavelength- and orientation-dependent study of the MOKE response presented in \Cref{fig:3_wavelength}.

The extracted odd Kerr rotation in saturation $\theta_\mathrm{odd}\left(H_\mathrm{sat}\right)$ attributed to the linMOKE is shown in \Cref{fig:3a,fig:3b} as a function of the crystallographic orientation $\left[hkl\right] \parallel \mathbf{H}_\mathrm{ext} \perp \hat{\mathbf{n}}$ and the probing wavelength $\lambda = 550\,\mathrm{nm} - 900\,\mathrm{nm}$. Concerning the crystallographic orientation $\left[hkl\right]$, an isotropic $\theta_\mathrm{odd}\left(H_\mathrm{sat}\right)$ is observed across all wavelengths in \Cref{fig:3a,fig:3b} which underlines the interpretation as the linear MOKE (first order contribution, with negligible higher order terms): Since the cubic crystal symmetry simplifies the first order magneto-optical tensor to a function of only one free parameter \supercite{visnovsky1986Magnetooptical, hamrlova2013Quadraticinmagnetization}, an isotropic first order contribution (linMOKE) as in \Cref{fig:3a,fig:3b} is expected for any wavelength. Yet, regarding the wavelength dependence of $\theta_\mathrm{odd}\left(H_\mathrm{sat}\right)$, a clear difference is observed for the two compounds: While \CMSi{} in \Cref{fig:3a} presents a significant wavelength dependence with a sign change at $\lambda \approx 600\,\mathrm{nm}$, \CMAl{} in \Cref{fig:3b} yields an approximately constant signal throughout the probed wavelength range $\lambda = 550\,\mathrm{nm} - 900\,\mathrm{nm}$. For a direct comparison, the directional average for each wavelength is presented in \Cref{fig:1c} in relation to the linMOKE rotation $\theta_\mathrm{linMOKE}$ reported on \CMSi{} by Silber et al. \supercite{silber2020Longitudinal, silber2020Scaling}. Hereby, we note that the reference data by Silber \textit{et al.} \supercite{silber2020Longitudinal} was recorded in a sophisticated in-depth study \supercite{silber2020Scaling} over a larger spectral range $\lambda = 225\,\mathrm{nm} - 1550\,\mathrm{nm}$ and with different angles of incidence. Moreover, as opposed to the $17\,\mathrm{nm}$ thick, epitaxial \CMSi{} film in our work grown by MBE directly on MgO and capped with Au, the reference by Silber et al. \supercite{silber2020Scaling} reports on a $30\,\mathrm{nm}$ thick \CMSi{} film grown by inductively coupled plasma-assisted magnetron sputtering on a Cr buffer layer, annealed at $500^\circ\mathrm{C}$, and capped with an Al layer \supercite{silber2020Scaling}. However, albeit these differences in setup and separation approach, as well as film growth and expected crystalline purity, and finally (magneto-)optical reflectivity of the different interfaces, the wavelength dependence of the \CMSi{} linMOKE data by Silber \textit{et al.} \supercite{silber2020Longitudinal, silber2020Scaling} is remarkably well reproduced in the present work after applying an empirical scaling factor of $-2$, where the sign reflects the opposite coordinate convention. The magnitude of the scaling factor may reflect the approximately double \CMSi{} layer thickness, as well as further differences in sample stacks and detection scheme between the two studies. We therefore restrict the comparison here to the spectral shape. 
We emphasise that this qualitative agreement not only highlights the reproducibility across different MOKE setups and differently grown \CMSi{} samples, but the confirmed comparability between the two studies allows to extend the existing reference data on \CMSi{} in \supercite{silber2020Scaling} by the additional \CMAl{} dataset reported in the present work. 

Turning to the extracted even Kerr rotation $\theta_\mathrm{even}$ attributed to the QMOKE, the signal in remanence $\theta_\mathrm{even}\left(0\right)$ is presented in \Cref{fig:3d,fig:3e} again normalised by the amplitude of the directly measured saturation signal $\theta_\mathrm{meas}\left(H_\mathrm{sat}\right)$ for an intuitive representation of the significance to the total measured MOKE signal. Contrary to the isotropic linMOKE in \Cref{fig:3a,fig:3b,fig:3c}, the directional dependence here presents the aforementioned sign change in the $\theta_\mathrm{even}\left(0\right)$ upon the crossing of a principal lattice axis $\langle100\rangle$. And similar to the linMOKE in \Cref{fig:3a,fig:3b,fig:3c}, a significant wavelength dependence is again only observed for the \CMSi{} film in \Cref{fig:3d}, whereas the \CMAl{} film yields an insignificant QMOKE contribution throughout the entire probed wavelength range $\lambda = 550\,\mathrm{nm} - 900\,\mathrm{nm}$ in \Cref{fig:3e}. 
For the wavelength dependence shown in \Cref{fig:3f}, the (non-normalised) QMOKE $\theta_\mathrm{even}\left(0\right)$ is extracted along the $[010]$ direction along which the effect is the strongest, as discussed around \Cref{fig:2_MOKE_anisotropy} above. The extracted data is again set in context with the reference data on \CMSi{} by Silber \textit{et al.} \supercite{silber2020Longitudinal, silber2020Scaling}, where the empirical scaling factor $4$ yields again a fair qualitative agreement in spectral shape with the reference data for \CMSi{}. Hereon, we remind that in this study $\theta_\mathrm{even}\left(0\right)$ is extracted at remanence, whereas Silber \textit{et al.}\supercite{silber2020Longitudinal, silber2020Scaling} obtain the QMOKE signal in saturation from measurements at four in-plane magnetisation directions. 
We note that even for this non-normalised QMOKE signal $|\theta_\mathrm{even}\left(0\right)|$, the extracted signal from the \CMAl{} film in \Cref{fig:3f} is nearly 0 for the entire probed wavelength range $\lambda = 550\,\mathrm{nm} - 900\,\mathrm{nm}$. The insignificant QMOKE signal in the \CMAl{} film is thus not only a perception due to the governing constant, non-vanishing linMOKE for all wavelengths, but also linked to a negligible second order contribution in this spectral range. 
This vanishing QMOKE might be related to the predominant B2 disorder in the \CMAl{} film, analogous to the aforementioned scaling of the QMOKE response with the L2\textsubscript{1} ordering degree in the case of \CMSi{} \supercite{wolf2011Quadratic, silber2020Scaling}. 

\FloatBarrier

\subsection{BLS signal intensities for various \texorpdfstring{Co$_2$Mn-based}{Co2Mn-based} Heusler compounds and relation to MOKE}

\begin{figure}[ht!]
  \centering
  \includegraphics{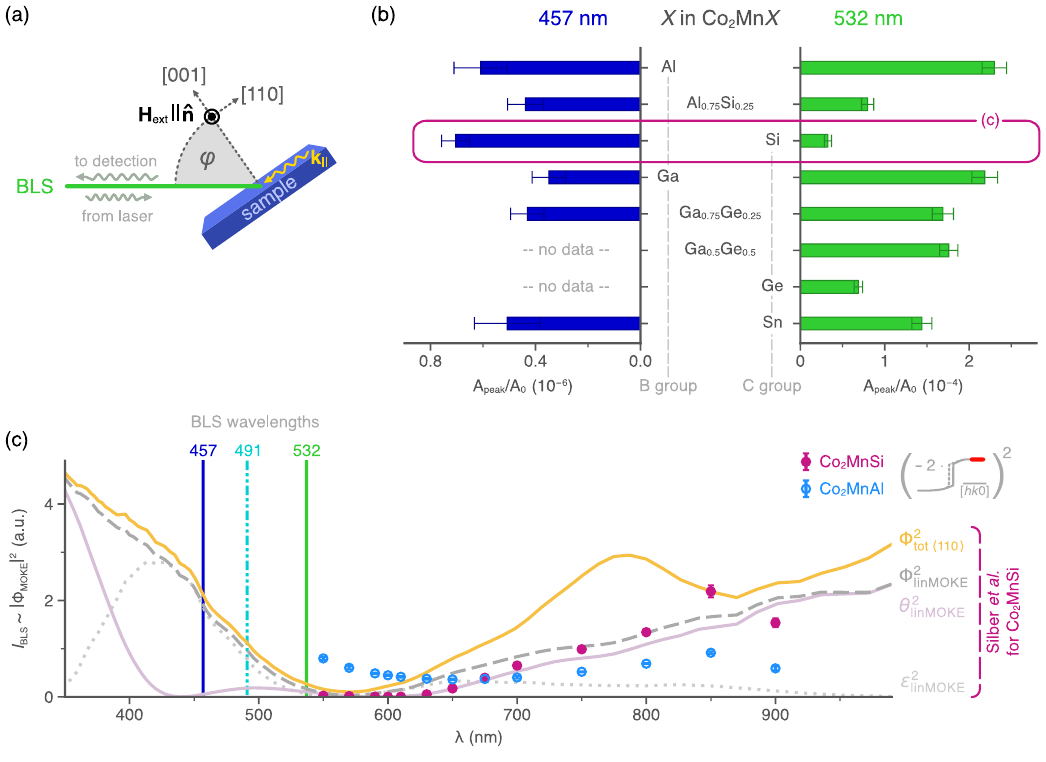}
  \caption[Correlation of MOKE and BLS signal]{
    \textbf{Correlation of MOKE and BLS signal.} 
    (a) Schematic of the BLS backscattering geometry. 
    (b) Normalised BLS peak amplitudes for the different \CMX{} compounds from BLS spectra of thermal spin waves measured for two wavelengths $\lambda_\mathrm{BLS} = 532\,\mathrm{nm}$ and $\lambda_\mathrm{BLS} = 457\,\mathrm{nm}$ with $\varphi_\mathrm{457\,\mathrm{nm}}=2.5^\circ$ and $\varphi_\mathrm{532\,\mathrm{nm}}=10^\circ$. 
    (c) Squared linear Kerr rotation $\theta^2_\mathrm{linMOKE}$ for \CMSi{} and \CMAl{} from \Cref{fig:3c}. Additionally, the spectral reference curves for \CMSi{} show the squared total Kerr angle $\Phi^2_\mathrm{tot\,\langle110\rangle}$, linear Kerr angle $\Phi^2_\mathrm{linMOKE}$ , linear Kerr rotation $\theta^2_\mathrm{linMOKE}$ and linear Kerr ellipticity $\varepsilon^2_\mathrm{linMOKE}$ from the reference study by Silber \textit{et al.} \supercite{silber2020Longitudinal, silber2020Scaling}. 
  }
  \label{fig:4_wavelength}

  \begin{subfigure}{0pt}
    \phantomcaption
    \label{fig:4a}
  \end{subfigure}
  \begin{subfigure}{0pt}
    \phantomcaption
    \label{fig:4b}
  \end{subfigure}
  \begin{subfigure}{0pt}
    \phantomcaption
    \label{fig:4c}
  \end{subfigure}
\end{figure}

The QMOKE anisotropy and the spectral dependence of the Kerr rotation particularly in \CMSi{} are not only of fundamental interest e.g. for the study of the crystal ordering, but of immediate practical relevance to the study of spin waves by Brillouin light scattering (BLS). 
The BLS signal intensity is analogously defined by the magneto-optical response\supercite{hamrle2010Analytical}, i.e. the Kerr amplitude of the dynamic magnetisation $\mathbf{m}$ reads
\begin{equation}
    I_\mathrm{BLS} \sim | \Phi_\mathrm{MOKE} \left(\mathbf{m} \right) |^2 = \theta_\mathrm{MOKE}^2 \left(\mathbf{m} \right) + \varepsilon_\mathrm{MOKE}^2 \left(\mathbf{m} \right)
    \label{eq:BLS_intensity}
\end{equation}
with the complex Kerr angle $\Phi_\mathrm{MOKE}\left(\mathbf{m} \right)$ composed of the Kerr rotation $\theta_\mathrm{MOKE}\left(\mathbf{m} \right)$ and Kerr ellipticity $\varepsilon_\mathrm{MOKE}\left(\mathbf{m} \right)$. The BLS probing geometry is illustrated in \Cref{fig:4a} showcasing the analogy to the MOKE probing above in \Cref{fig:1a}. We note that as opposed to the MOKE setup with $\mathbf{H}_\mathrm{ext} \perp \hat{\mathbf{n}}$, the external field is applied $\mathbf{H}_\mathrm{ext} \parallel \hat{\mathbf{n}}$ in the BLS study, but still corresponds to the same symmetry family as coinciding with the incidence plane of light. 
In relation, the static Kerr spectra can serve as a proxy for the spectral trend of the dynamic BLS response, assuming (i) dynamics in the linear regime, i.e. small dynamic magnetisation $\mathbf{m}$ (valid for the weak thermal excitation), and (ii) static and dynamic magneto-optical response governed by the same magneto-optical tensor (valid for GHz dynamics $\ll$ optical frequencies). Moreover, the comparison across the sample series is based on the assumption of similar mode profile and depth sensitivity across the sample series (valid for this \CMX{} sample series with comparable film stacks).

\pagebreak[4]

\Cref{fig:4b} presents the BLS signal intensities for the \CMX{} (\textit{X} = \{Al$_x$Si$_{1-x}$, Ga$_x$Ge$_{1-x}$, Sn\}) sample series extracted from BLS spectra of thermal spin waves for two laser wavelengths $\lambda_\mathrm{BLS} = 457\,\mathrm{nm}$ and $\lambda_\mathrm{BLS} = 532\,\mathrm{nm}$. We note that the normalised intensities must not be compared across the two wavelengths\footnote{The normalised amplitudes at the two wavelengths differ by more than two orders of magnitude (note the different axis prefactors, $10^{-6}$ vs. $10^{-4}$) and cannot be compared directly: Changing $\lambda_\mathrm{BLS}$ requires exchanging the laser and realigning the setup, which can result in a different laser power on the sample. In addition, optical components such as beam splitters are optimised for specific wavelengths and the detector sensitivity is wavelength-dependent, as are the reflectivity and absorption of the Heusler film as well as the Au capping layer. The normalisation to the reference beam removes laser-power drifts \emph{within} a measurement, but not these cross-wavelength factors. Note that the inelastically scattered light intensity scales with the sample reflectivity\supercite{hamrle2010Analytical}, and the normalisation on the reference beam cannot remove such optical reflectivity differences between the samples at one given wavelength. We assume comparable optical reflectivities at a given wavelength $\lambda_\mathrm{BLS}$ for the thin film series of similar metallic \CMX{} compounds in analogous film stacks and all with the same Au capping.} and therefore restrict the interpretation to the relative differences between compounds measured under identical conditions. Two features stand out in \Cref{fig:4b}: 
Firstly, a wavelength-dependent clustering of \CMX{} into Boron-group (B group) elements \textit{X} = \{Al, Ga\} and Carbon-group (C group) elements \textit{X} = \{Si, Ge, Sn\} is observed. For $\lambda_\mathrm{BLS} = 532\,\mathrm{nm}$ the magneto-optical signal intensity shows a clear correlation of highest (lowest) signal intensities with the Boron (Carbon) group elements, not only comparing the different ternary compounds \CMX{} with \textit{X} = \{Al, Si, Ga, Ge, Sn\} but also within the quaternary series \CMAS{} and \CMGxGx{}. For $\lambda_\mathrm{BLS} = 457\,\mathrm{nm}$ on the other hand, no clear correlation with the elemental group is identified. The Boron- and Carbon-group elements differ by one valence electron, and while this also manifests in the associated Slater-Pauling behaviour clustering the magnetic moment by the number of valence electrons, the spectral dependence of the elemental correlation points to a band-filling origin. 
Secondly, \CMSi{} displays by far the strongest wavelength dependence in relation to the other compounds in the series. For $\lambda_\mathrm{BLS}=532\,\mathrm{nm}$ it yields by far the weakest signal of the series, whereas it recovers to among the strongest signals for $\lambda_\mathrm{BLS} = 457\,\mathrm{nm}$. This wavelength dependence tracks in the complex Kerr angle in \Cref{fig:4c}, where the measured $\theta_\mathrm{linMOKE}^2$ (this work) and the reference $|\Phi|^2$ spectra by Silber \textit{et al.}\supercite{silber2020Scaling, silber2020Longitudinal} for the $\langle 110\rangle$ BLS geometry ($\Phi_\mathrm{tot\,\langle 110\rangle}^2$) collapse to a minimum around $\lambda \approx 550\,\mathrm{nm} - 620\,\mathrm{nm}$ and recover towards the blue-light regime $\lambda \lesssim 500\,\mathrm{nm}$. In comparison, \CMAl{} shows a finite complex Kerr rotation with negligible wavelength dependence throughout the entire probed regime $\lambda = 550\,\mathrm{nm} - 900\,\mathrm{nm}$, yet the data is insufficient to directly correlate it with the BLS response both because of the non-overlapping wavelength ranges and because only the Kerr rotation $\theta_\mathrm{linMOKE}$ was measured: The reference curves in \Cref{fig:4c} show that the blue-side recovery for \CMSi{} is carried mainly by the linear Kerr ellipticity $\varepsilon_\mathrm{linMOKE}^2$ rather than the rotation $\theta_\mathrm{linMOKE}^2$, underlining that the BLS-relevant quantity is the full Kerr angle $I_\mathrm{BLS}\sim|\Phi_\mathrm{MOKE}|^2$ in \Cref{eq:BLS_intensity}, not the Kerr rotation alone. Both points, i.e. the wavelength-dependent band-filling sensitivity as well as the composition-dependent relative change between wavelengths, connect back to the aforementioned open question on the key factors for the higher order magneto-optical effects: Notably, \CMGa{} shares the full L2\textsubscript{1} ordering of \CMSi{} but has a different number of valence electrons and moreover its band structure features a near-vanishing spin band gap $\sim0\,\mathrm{eV}$ as opposed to the $0.7\,\mathrm{eV}$ in the \CMSi{} film \supercite{guillemard2020Issues}. Yet a conclusive analysis would require additional spectroscopic data. 
Nevertheless, for \CMSi{}, \Cref{fig:4c} highlights the agreement between the measured MOKE spectra in this work for $\lambda = 550\,\mathrm{nm} - 900\,\mathrm{nm}$ and the reference MOKE spectra from Silber \textit{et al.}\supercite{silber2020Longitudinal, silber2020Scaling}, the spectral dependence of which is reflected in the matching BLS intensity trend. 
This is consistent with the analytical MOKE-BLS link\supercite{hamrle2010Analytical}, which is further corroborated by a recent successful study using the intermediate wavelength $\lambda = 491\,\mathrm{nm}$ \supercite{friedel2026Epitaxial}, marked by the dash-dotted line in \Cref{fig:4c}. These results show that MOKE spectroscopy can be a practical guide for the choice of BLS probing wavelength. 

Ultimately, we note that here the BLS study was carried out along a $\langle110\rangle$ axis with negligible QMOKE contribution at the two probed BLS wavelengths (compare the reference curves for $| \Phi_\mathrm{tot \langle110\rangle} |^2$ and $| \Phi_\mathrm{linMOKE} |^2$ in \Cref{fig:4c}), and emphasise that this does not hold in general: A negligible QMOKE signal at a given geometry and wavelength clearly cannot allow to exclude a potentially significant QMOKE to the dynamic response in other configurations, given the strong wavelength dependence as well as directional dependence of the second-order effect established above, with a particular significance for a probing close to the high symmetry directions $\langle100\rangle$. Importantly, we emphasise that the direct link of MOKE and BLS intensity is not only relevant as a guide for the choice of BLS probing wavelength, but may also require consideration in the subsequent analysis of the probed spin-wave dynamics: 
The higher-order magneto-optical response is nonlinear in the magnetisation, and for the second-order effect specifically mixes the longitudinal and transverse components in the aforementioned terms $\theta_{M_\mathrm{L} M_\mathrm{T}} \sim M_\mathrm{L} M_\mathrm{T}$ and $\theta_{M_\mathrm{L}^2 - M_\mathrm{T}^2} \sim M_\mathrm{L}^2 - M_\mathrm{T}^2$ (analogous to the planar Hall effect and anisotropic magneto-resistance \supercite{luo2026SymmetryDriven}). Consequently, a non-negligible higher order magneto-optical contribution could mix harmonic responses of a mode oscillating at $\omega$ and potentially give rise to a magneto-optical artifact at higher harmonic frequencies in the BLS detection as opposed to genuinely populated higher harmonic modes. Although this is negligible for thermally excited magnons as studied here, particular attention is suggested for large-amplitude driven modes studied in nonlinear magnonics especially in view of the growing interest in higher harmonic generation\supercite{koerner2022Frequency, dreyer2022Imaging, lendinez2023Nonlinear, nikolaev2024Resonant} and the recent experimental reports on significant higher-order magneto-optic effects \supercite{gaerner2024Cubic, silber2026Cubicinmagnetizationa, gaerner2026Cubic}. 

\FloatBarrier

\section{Conclusions}
\label{sec:conclusions}

We have studied the magneto-optical response of the epitaxial \CMX{} (\textit{X} = \{Al$_x$Si$_{1-x}$, Ga$_x$Ge$_{1-x}$, Sn\}) Heusler series in a combined investigation using MOKE spectroscopy as well as BLS spectroscopy of thermally populated magnons. Angle-resolved MOKE measurements at $\lambda = 636\,\mathrm{nm}$ identify a significant quadratic MOKE contribution only in \CMSi{}, maximal along the $\langle100\rangle$ axes and reaching up to $\sim$150\% of the total signal, whereas \CMAl{}, \CMGa{} and \CMSn{} respond dominantly linearly. In the wavelength-dependent study across $\lambda = 550\,\mathrm{nm}-900\,\mathrm{nm}$ the extracted Kerr rotation of \CMSi{} is strongly dispersive and reproduces the reference spectra of Silber \textit{et al.} \supercite{silber2020Scaling, silber2020Longitudinal}. In comparison, the extracted Kerr rotation of \CMAl{} remains nearly constant with a finite first order contribution and vanishing higher order effects throughout the entire probed range $\lambda = 550\,\mathrm{nm}-900\,\mathrm{nm}$. 

These magneto-optical properties directly govern the BLS signal: \CMSi{} yields the weakest intensity of the series at the common $\lambda_\mathrm{BLS} = 532\,\mathrm{nm}$ and recovers to among the strongest at $\lambda_\mathrm{BLS} = 457\,\mathrm{nm}$, tracking the spectral dependence of $|\Phi_\mathrm{MOKE}|^2$. Additionally, we find that the BLS intensities at $\lambda_\mathrm{BLS} = 532\,\mathrm{nm}$ cluster by the valence-electron count of the \textit{X} element in \CMX{} (Boron vs. Carbon group), pointing to a band-filling origin. Concerning this, our study identifies two interesting routes for follow up work to further narrow down the intrinsic origins of the higher order magneto-optical response: Firstly, a full spectroscopic study of the MOKE response  on the \CMSn{} compound would give insight into the influence of the inverse X order compared to the full L2\textsubscript{1} order. And secondly, a full spectroscopic study of the MOKE response on \CMGa{} in relation to the existing data on \CMSi{} could allow for a comparison between L2\textsubscript{1} ordered materials with different valence electron count and band gap.

Finally, our results provide experimental support for the analytical link between MOKE and BLS\supercite{hamrle2010Analytical} and establish MOKE spectroscopy as a practical guide for selecting the BLS probing wavelength. 
We emphasise that this general conclusion is not restricted to Heusler compounds, but applicable to any other magnetic material. We note that it is generally advisable to record the Kerr spectra on the same sample as the BLS measurements, since the magneto-optical response generally depends on the structural state, as is for example the case in this study for the chemical ordering in the \CMX{} compounds. Beyond guiding the wavelength choice, the direct link between MOKE and BLS calls for attention in the analysis of driven spin-wave dynamics in systems with non-negligible higher order magneto-optical response, where the higher order contribution may yield a magneto-optical detection artifact mimicking genuinely populated higher-harmonic magnons.

\section*{Acknowledgements}
The authors thank the European Research Council (ERC) for funding this work through the ERC Starting Grant 101042439 (CoSpiN), as well as the Agence Nationale de la Recherche (France) for funding this work via the contracts ANR-20-CE24-0012 (MARIN) and ANR-20-CE24-0023 (CONTRABASS). 
A.M. Friedel acknowledges financial support from the Franco-German University (FGU) via the cotutelle de thèse stipend CT-15-25. 
The funders played no role in study design, data collection, analysis and interpretation of data, or the writing of this manuscript. 
A.M. Friedel thanks Lukas Körber for valuable advice on data visualisation and figure design. The authors thank Timo Kuschel for helpful comments and suggestions on the first version of this manuscript. 

\section*{Author contributions}
A.M.F. and P.P. conceived the study. N.F. carried out the MOKE measurements as well as the BLS measurements at $457\,\mathrm{nm}$ under the supervision of A.M.F. and P.P., T.B. carried out the BLS measurements at $532\,\mathrm{nm}$. A.M.F., N.F. and T.B. analysed the data. A.M.F. visualised the results. All authors discussed the results. P.P., S.P.-W. and S.A. acquired funding. A.M.F. wrote the original manuscript. All authors reviewed and commented on the manuscript. 

\section*{Data availability statement}
The datasets generated and analysed during this study are available in the ZENODO repository: \url{https://doi.org/10.5281/zenodo.21323893}.

\printbibliography

\end{document}